\documentclass[11pt]{article}
\usepackage{graphicx,amsmath,bm, amsthm,mathrsfs,amssymb,braket, verbatim}
\usepackage{caption}
\usepackage{subcaption}
\usepackage[usenames]{color}
\usepackage{ulem,mathtools}
\usepackage{pdfpages}
\usepackage{lscape}
\usepackage{relsize}
\usepackage{authblk}
\usepackage{cite}
\usepackage{tcolorbox}
\usepackage[section]{placeins}
\usepackage{placeins}

\newcommand{\ee}{\end{equation}}

\newcommand{\reff}[1]{(\ref{#1})}
\newcommand{\beq}{\begin{equation}}
\newcommand{\eeq}[1]{\label{#1}\end{equation}}
\newcommand{\beqa}{\begin{eqnarray}}
\newcommand{\eea}{\end{eqnarray}}
\newcommand{\eeqa}[1]{\label{#1}\end{eqnarray}}
\newcommand{\beg}{\begin{equation*}}
\newcommand{\eeg}{\end{equation*}}

\newcommand{\bsplit}{\begin{split}}
\newcommand{\esplit}{\end{split}}

\newcommand\blue[1]{{\textcolor{blue}{#1}}} 
\usepackage{circuitikz} %To draw the electrical circuits
\usepackage[capposition=bottom]{floatrow} %To add notes under a figure
\usepackage{epigraph} %For quotes at the beginning of sections and text
\usepackage{appendix}
\usepackage{rotating}
\allowdisplaybreaks

\title{Finite path integral limits work \\in cases where the perturbative series\\ is not Borel summable}
\author[]{Ariel Edery\thanks{aedery@ubishops.ca}}
\affil[]{Department of Physics and Astronomy, Bishop's University, 2600 College Street, Sherbrooke, Qu\'{e}bec, Canada, J1M 1Z7.\vspace{1em}}
\begin{document}
\date{}
\maketitle
\begin{abstract}
The perturbative expansion in powers of the coupling of observables in quantum field theory and quantum mechanics is known to yield an asymptotic series. If the original physical system is well-behaved and a finite observable is expected, this can often be calculated via a Borel resummation of the asymptotic series. However, there are cases where a system is well-behaved and the series is not Borel summable. This typically occurs when the physical system has a non-trivial vacuum structure. It has recently been shown that if the perturbative series is carried out under finite path integral limits, one can obtain a convergent series that yields observables even at strong coupling. This was recently used to obtain the energy at strong coupling for the anharmonic oscillator. This is a Borel summable case so the question is whether finite path integral limits work when the series is not Borel summable. To begin answering this question we consider a simple non-Borel summable case: the series stemming from a basic integral where the function has a double-well shape and hence two minima. The integral has an exact analytical expression that the series can be compared to. Under finite integral limits that run from $-L$ to $L$, where $L$ is finite, positive and real, we develop two perturbative series in powers of the coupling: one by expanding the integral about the local maximum at the origin and the other by expanding it about one of the minima. In both cases, we obtain an absolutely convergent series and the series sums to the exact analytical expression of the original integral in the infinite $L$ limit. It is significant that a perturbative  expansion about one of the minima reproduces the exact analytical expression because this implies that it captures the full effect of both minima. 
\end{abstract}
\setcounter{page}{1}
\newpage
\section{Introduction}\label{Intro}
It is well known that the standard perturbative expansion in powers of the coupling for observables in quantum field theory (QFT) and quantum mechanics yields an asymptotic series \cite{Dyson, Zinn, McKane}. In some cases, a finite value can be extracted from the asymptotic series via Borel resummation \cite{Marino1,Strocchi}. However, there are cases where the physical system is well behaved with finite observables and yet the asymptotic series is not Borel summable. Examples include the energy series for the double-well potential in quantum mechanics \cite{Marino1} and infrared renormalons in Quantum Chromodynamics \cite{Ioffe,Beneke}. This usually occurs because the system has a non-trivial vacuum as in the above two cases \cite{Zinn,Ioffe}. Recently, it has been identified that the asymptotic nature of the perturbative series of an otherwise finite original integral is caused by evaluating the series under infinite limits of the path integral \cite{Ariel1,Ariel2}. By evaluating the series under finite path integral limits, it has been shown that one can obtain a convergent perturbative series and extract from it observables even at strong coupling. This was recently used to obtain the energy at strong coupling of the anharmonic oscillator via a convergent perturbative energy series in powers of the coupling \cite{Ariel2}. It is well-known that the asymptotic energy series of the anharmonic oscillator \cite{BenderWu1, BenderWu2, BenderWu3} is also Borel summable \cite{Graffi,Marino1, Ioffe}. So an important question is whether finite path integral limits work in cases where the series is not Borel summable. In this work, we begin to answer this question by considering a basic integral where the integrand is the exponential of a function with a double-well shape and hence has two minima. This simple example contains the two essential features without the technicalities: it has a 'non-trivial vacuum' (i.e. the two minima) and its standard perturbative series in powers of the coupling $\lambda$ yields an asymptotic series which is not Borel summable. Another advantage is that the integral yields an exact analytical expression valid for all values of the coupling; this is very useful as we have a solid benchmark to compare the series to. Using finite integral limits, we obtain two convergent perturbative series in powers of the coupling: one is based on an expansion about the local maximum at the origin and the second is based on an expansion about one of the minima. Both series converge for all values of the coupling $\lambda$ and we show that in the infinite integral limit, the sum of the series yields the exact analytical expression of the original integral. This is an important result because even when the perturbative series is expanded about one of the two minima, it captures the effects of both minima that appears in the full non-perturbative result i.e. the analytical expression. 

The paper is organized as follows. In section 2 we obtain the standard perturbative asymptotic series for the basic integral by expanding about one of the minima and show that it is not Borel summable. In section 3 we develop the convergent perturbative series under finite path integral limits and show that in the infinite integral limit they reproduce the exact analytical expression for the integral. In section 3.1 the series is obtained by expanding about the local maximum at the origin $x=0$ and in section 3.2 the series is obtained by expanding about one of the minima. Section 4 is the conclusion where we discuss the results and future work.    
            
\section{The non-Borel summable case: a function with two minima}\label{InfLimits}

Consider the integral
\beq 
I= \int_{-\infty}^{\infty} e^{\frac{1}{2} a\,x^2 -\lambda \,x^4} \,dx
\eeq{I1}
where $a>0$ and $\lambda>0$. This integral is finite and can be expressed analytically in terms of modified Bessel functions of the first kind $I_n(z)$:
\beq
I_{analytic}=\frac{\pi}{4} \,\, e^{\frac{a^2}{32 \lambda }} \,\sqrt{\frac{a}{\lambda }} \,\,\Big[\,I_{\frac{1}{4}}\big(\tfrac{a^2}{32 \lambda }\big)+I_{\frac{-1}{4}}\big(\tfrac{a^2}{32 \lambda }\big)\,\Big]\,.
\eeq{I2}
The integrand in \reff{I1} is $e^{-f(x)}$ where $f(x) =\lambda \,x^4-\tfrac{1}{2} a \,x^2$. This function has a double-well shape and is symmetric under reflections $x\to-x$. It has two minima located at $x=x_{\pm}= \pm \sqrt{\tfrac{a}{4\,\lambda}}$ and a local maximum at $x=0$. 
Let us now expand the original integral \reff{I1} as a power series in $\lambda$. If we try to expand $e^{-\lambda \,x^4}$ in a power series about $\lambda=0$ and multiply this by $e^{a\,x^2/2}$, each term  yields infinity after integration since $a$ is positive. The integral of $e^{a\,x^2/2}$, the quadratic part, is not a Gaussian here since $a$ is positive and therefore diverges exponentially. Note that expanding the term $e^{-\lambda \,x^4}$ about $\lambda=0$ yields the same series as expanding it about $x=0$. This means we cannot expand the series about the origin which is a local maximum. We therefore expand the integrand about one of the minima. It does not matter which minima we choose. We expand about the positive minima $x_{+}$ by writing  $x=\sqrt{\tfrac{a}{4\,\lambda}} +y$. The integrand is now expressed in terms of the variable $y$. After this substitution, we can obtain a series in powers of $\lambda$ by expanding about $\lambda=0$ which is equivalent to expanding the non-quadratic part about $y=0$. The series expansion of integral \reff{I1} is then given by 
\begin{align}
I_{series}&=e^{\tfrac{a^2}{16\,\lambda}} \int_{-\infty}^{\infty} e^{-a\, y^2 -2 \sqrt{a}\,\sqrt{\lambda }\, y^3-\lambda \, y^4} \,dy\nonumber\\
&=e^{\tfrac{a^2}{16\,\lambda}} \int_{-\infty}^{\infty} e^{-a\, y^2} \sum_{n=0}^{\infty}\frac{(-1)^n}{n!}\, (\,2\, \sqrt{a\,\lambda}\, y^3+\lambda \, y^4\,)^n\,dy\,.
\label{IS}
\end{align}
Before integrating, we perform a binomial expansion of the quantity inside the sum 
\begin{align}
\frac{(-1)^n}{n!}\, (\,2\,\sqrt{a\,\lambda}\, y^3+\lambda \, y^4\,)^n&=\sum_{k=0}^n \frac{(-1)^n}{n!} \,\binom{n}{k}\,(\,2\, \sqrt{a\,\lambda}\, y^3\,)^{n-k}\,\,(\lambda\,y^4\,)^k\nonumber\\
&=\sum_{k=0}^n (-1)^n\, \frac{2^{n-k}}{k!\, (n-k)!}\,\,a^{\tfrac{n-k}{2}}\,\,\lambda^{\tfrac{n+k}{2}}\,y^{3n+k}\,.
\label{Binom}
\end{align}
The integral over $y$ yields:
\beq
\int_{-\infty}^{\infty} e^{-a\, y^2}\,y^{3n+k}\, dy= \frac{1}{2}(1+(-1)^{n+k})\,a^{-\tfrac{(3n+1+k)}{2}}\,\Gamma\big(\tfrac{3n+1+k}{2}\big)\,.
\eeq{Inty}
The above integral is zero unless $n+k$ is even. Substituting the above result as well as \reff{Binom} into \reff{IS} yields
\beq
I_{series}=e^{\tfrac{a^2}{16\,\lambda}}\,\sum_{n=0}^{\infty} \sum_{k=0}^n (-1)^n \frac{2^{n-k}}{k!\, (n-k)!}\,a^{-n-k-\tfrac{1}{2}}\,\tfrac{1}{2}\big(1+(-1)^{n+k}\,\big)\,\Gamma\big(\tfrac{3n+1+k}{2}\big)\,\,\lambda^{\tfrac{n+k}{2}}\,.
\eeq{IS2}
Only terms in the sum where $n+k$ is even survive. Since the power of $\lambda$ is $(n+k)/2$ this implies that the series contains only positive \textit{integer} powers of $\lambda$ (i.e. no half-integers). We want to organize the series order by order in powers of $\lambda$. Let $q$ represent the integer powers of $\lambda$. Then the set of values $(n,k)$ that satisfy $(n+k)/2=q$ are $(2\,q-k,k)$ with $k$ running from $0$ to $q$ inclusively i.e. the set $\{(2\,q,0)\,,\,(2\,q - 1,1)\,,\,(2\,q-2,2)\,,...\,,\,(q,q)\,\}$. We can therefore set $n=2\,q-k$ in \reff{IS2} and have $k$ run from $0$ to $q$ with $q$ running from $0$ to $\infty$. This yields
\begin{align} 
I_{series}&=e^{\tfrac{a^2}{16\,\lambda}}\,\sum_{q=0}^{\infty} \sum_{k=0}^q (-1)^k \frac{4^{q-k}}{k!\, (2q-2k)!}\,a^{-2q-\tfrac{1}{2}}\,\Gamma\big(3q-k+1/2\big)\,\,\lambda^q\nonumber\\
&=e^{\tfrac{a^2}{16\,\lambda}}\,\sum_{q=0}^{\infty} \frac{4^q\,\lambda^q}{a^{2q+\tfrac{1}{2}}}\,\sum_{k=0}^q \frac{(-1)^k\,\Gamma\big(3q-k+1/2\big)}{4^k\,k!\, (2q-2k)!}\,.
\label{IS3} 
\end{align} 
The sum over $k$ can be readily evaluated and yields 
\beq
\sum_{k=0}^q \frac{(-1)^k\,\Gamma\big(3q-k+1/2\big)}{4^k\,k!\, (2q-2k)!}=\frac{\Gamma(2q+1/2)}{q!}= \frac{(4q-1)!!}{4^q\,q!}\,\sqrt{\pi}
\eeq{Sumq}
where the double factorial is defined in the standard fashion. Substituting the above into \reff{IS3} we finally obtain our series:
\begin{align}
I_{series}&=e^{\tfrac{a^2}{16\,\lambda}}\,\sum_{n=0}^{\infty} \frac{4^n}{a^{2n+\tfrac{1}{2}}}\frac{(4n-1)!!}{4^n\,n!}\,\sqrt{\pi}\,\lambda^n\nonumber\\
&=e^{\tfrac{a^2}{16\,\lambda}}\,\sqrt{\frac{\pi}{a}}\,\sum_{n=0}^{\infty}\frac{(4n-1)!!}{a^{2n}\,n!}\,\lambda^n\nonumber\\
&=e^{\frac{a^2}{16 \lambda }}\, \sqrt{\frac{\pi}{a}}\,\sum_{n=0}^{\infty} C_n\,\lambda^n
\label{IS4}
\end{align}
where the coefficients $C_n$ are given by
\beq
C_n=\frac{(4n-1)!!}{a^{2n}\,n!}\,.
\eeq{Cn}
We write out below the first few terms of the series:
\beq 
I_{series}= e^{\frac{a^2}{16 \lambda }}\, \sqrt{\frac{\pi}{a}}\, \left(1+\frac{3\, \lambda }{a^2}+\frac{105 \,\lambda ^2}{2\, a^4}+\frac{3465 \,\lambda ^3}{2\, a^6}+\frac{675675 \,\lambda ^4}{8 \,a^8}+\ldots\right)\,.
\eeq{Terms}
Note that $C_n$ is always positive so that each term in the series has the same sign. It is clear that the series \reff{IS4} is an asymptotic (divergent) series since $\lim_{n\to \infty} \,C_n\to \infty$. An expansion of $C_n$ about large (infinite) order $n$ yields 
\beq
C_{n_{{}_{\,\,(n\to \infty)}}}\to \Big(\frac{16}{a^2}\Big)^n \,e^{-n}\, \frac{n^n}{\sqrt{n\,\pi}}\approx \Big(\frac{16}{a^2}\Big)^n\,\frac{(n-1)!}{\sqrt{2}\,\pi} 
\eeq{CnS}
where the approximation sign stems from applying Stirling's approximation. $C_n$ grows factorially and has the form $A^{-n} \,(n-1)!$ where $A=a^2/16$. This value will show up again later below. The fact that $A$ is positive is in accord with our previous statement that the terms in the series have the same sign. We are now going to show that series is \textit{not} Borel summable. The prefactor in $I_{series}$ plays no role in proving this and we can simply focus on the series $S(\lambda)=\sum_{n=0}^{\infty} C_n\,\lambda^n$. The Borel transform of $S(\lambda)$ is given by
\beq
\hat{S}(b)=\sum_{n=0}^{\infty} \frac{C_n}{n!}\, b^n= \sum_{n=0}^{\infty} \frac{(4n-1)!! }{ a^{2 n}\,(n!)^2}\, b^n =\frac{2 K\left(\frac{8 \sqrt{b}}{a+4 \sqrt{b}}\right)}{\pi \,\sqrt{\frac{a+4 \sqrt{b}}{a}}}
\eeq{BorelT}
where $K(m)$ is the complete elliptic integral of the first kind. It is defined as
\beq
K(m)=\int_0^{\pi/2} \frac{d\theta}{\sqrt{1-m\, \sin^2\theta}}
\eeq{Km}
which has a logarithmic singularity at $m=1$. Therefore the Borel transform $\hat{S}(b)$ has a logarithmic singularity when $\frac{8 \sqrt{b}}{a+4 \sqrt{b}}=1$ which occurs at $b=a^2/16$. This is equal to the value of $A$ we found above. We can readily see why. The large order behaviour of $C_n$ is proportional to $A^{-n} \,(n-1)!$. Substituting this into the Borel transform yields $\sum_{n=n_0}^{\infty} (b/A)^n/n =$finite terms +$ \log \left(\frac{A}{A-b}\right)$. We clearly see that there is a logarithmic singularity when $b=A=a^2/16$. The Borel resummation is defined by the integral 
$\tfrac{1}{\lambda} \int_0^\infty e^{-b/\lambda} \,\hat{S}(b)\,db$. This cannot be calculated since the singularity is on the \textit{positive real axis} along the path of integration. The series $S(\lambda)$ is therefore not Borel summable which implies that the asymptotic series $I_{series}$ given by \reff{IS4} cannot be resummed to yield the finite analytical result \reff{I2} of the original integral \reff{I1}.  

Note that the non-trivial vacuum structure of $f(x)$, which has the two degenerate minima (a double-well shape), is what is ultimately responsible for the fact that the series is not Borel summable. In particular, we could not expand $e^{-\lambda \,x^4}$ about the local maximum at $x=0$. Instead, we had to expand the series about one of the two minima. This led to a series where the terms had the same sign so that the singularity of the Borel transform was on the positive real axis. The singularity was therefore along the path of integration of the Borel resummation and hence not Borel summable.  In contrast, the anharmonic oscillator \cite{Marino1} has a unique minimum at $x=0$. An expansion about $x=0$ of $e^{-\lambda \,x^4}$ leads to an asymptotic series where the terms have alternating signs. The singularity of the Borel transform is then negative. This lies outside the path of integration of the Borel resummation and hence is Borel summable. Borel summability is closely related to the asymptotic series defining a unique function \cite{Zinn}.   

\section{Finite path integral limits and the non-Borel summable case}

We will now show that using finite path integral limits one can develop a perturbative series in powers of $\lambda$ which is absolutely convergent for the non-Borel summable case of the previous section. We will do this in two ways: by expanding the series about the local maximum at the origin $x=0$ and about one of the minima. Moreover, we will show in both cases that the absolutely convergent series can be summed to yield in the infinite integral limit the analytical expression \reff{I2} of the original integral \reff{I1}. Recall that in the previous section we were not able to perform an expansion about $x=0$ since the infinite integral limits meant each term in the series would be infinite. With finite integral limits, the expansion about $x=0$ does not have this issue.    

\subsection{Expansion about the local maximum at $x=0$}

We begin by rewriting the original integral \reff{I1} in the following equivalent fashion:
\begin{align}
I &=\int_{-\infty}^{\infty} e^{\frac{1}{2} a\,x^2 -\lambda \,x^4} \,dx\nonumber\\ 
&=\lim_{L\to \infty}\int_{-L}^L  e^{\frac{1}{2} a\,x^2 -\lambda \,x^4} \,dx\nonumber\\
&=\lim_{L\to \infty}\,I(L)
\label{IL}
\end{align}
where 
\beq
I(L)=\int_{-L}^L  e^{\frac{1}{2} a\,x^2 -\lambda \,x^4}\,dx.
\eeq{IL2}
Here $L$ is finite, positive and real. Note that in \reff{IL} we have not altered the original integral in any way. The infinite limits which originally appear on the integral sign have simply been written by placing finite limits $L$ on the integral with the limit of $L$ taken to infinity. In fact, integrals with infinite limits are usually defined this way. We can now perform the perturbative series expansion in powers of $\lambda$ on $I(L)$ which has finite integral limits. This makes an important difference as this yields an absolutely convergent series instead of an asymptotic one \cite{Ariel1, Ariel2}. There will therefore be no need for any resummation procedure (Borel or otherwise). Though $L$ is finite it can be made arbitrarily large so that the series can be evaluated to any desired level of accuracy. If the summation of the infinite series yields an analytical expression as a function of $L$, one can also take the infinite $L$ limit of the expression to obtain the exact answer. This is precisely what we will do here. 

The series epansion of $I(L)$ in powers of $\lambda$ can be obtained by simply expanding $e^{-\lambda \,x^4}$ about $\lambda=0$ in the integrand. This is equivalent to expanding about $x=0$. This yields the series
\begin{align}
I_{series}(L)&= \int_{-L}^L dx\, e^{\frac{1}{2} a\,x^2}\sum_{n=0}^{\infty} \frac{(-\lambda \,x^4)^n}{n!}
\nonumber\\
&= \sum_{n=0}^{\infty} \frac{(-1)^n}{n!} \,\lambda^n \int_{-L}^L dx \,e^{\frac{1}{2} a\,x^2} \,x^{4n} \nonumber\\
&=\sum_{n=0}^{\infty} \frac{(-1)^n}{n!} \,\lambda^n \Big(-i\,\gamma\,\big(\,2n +\tfrac{1}{2},-a\,L^2/2\,\big)\, \big(\tfrac{2}{a}\big)^{2n+\tfrac{1}{2}}\,\Big)\nonumber\\
&=\sum_{n=0}^{\infty} b_n \,\lambda^n
\label{ILS}
\end{align}
where the coefficients $b_n$ are real and given by
\beq
b_n=\frac{(-1)^{n}}{n!}\Big(-i\,\gamma\,\big(\,2n +\tfrac{1}{2},-a\,L^2/2\,\big)\, \big(\tfrac{2}{a}\big)^{2n+\tfrac{1}{2}}\,\Big)
\eeq{bn}
and $\gamma(a,y)$ is the incomplete gamma function defined by 
\beq
\gamma(a,y)=\int_0^y t^{a-1}\,e^{-t}\,dt\,.
\eeq{GG}
Since $-a\,L^2/2$ is negative, it follows that $-i\,\gamma\,\big(\,2n +\tfrac{1}{2},-a\,L^2/2\,\big)$ is positive and real (as expected since this quantity stems from a positive integral). 

The series \reff{ILS} is an absolutely convergent series since
\begin{align}
&\lim_{n\to \infty} \Big|\frac{b_{n+1}\,\lambda^{n+1}}{b_{n}\,\lambda^{n}}\Big|\nonumber\\
= &\lim_{n\to \infty} \Big|\lambda\,\Big(\frac{2}{a}\Big)^{1/2} \,\frac{\gamma\,\big(\,2n +\tfrac{3}{2},-a\,L^2/2\,\big)}{(n+1)\,\gamma\,\big(\,2n +\tfrac{1}{2},-a\,L^2/2\,\big)}\Big|\nonumber\\
=&\,\,0\,.
\end{align}    
Since the above limit is zero, and hence less than unity, by the ratio test the series $I_{series}(L)$ given by \reff{ILS} is an absolutely convergent series in powers of $\lambda$.  Note that this convergence holds regardless of the value of the coupling constant $\lambda$ (i.e. valid for weak and strong coupling). We now show explicitly the first few terms of the series $I_{series}(L)$ to obtain a sense of its structure:  
\begin{align}
I_{series}(L)&= -i\,\gamma\,\big(1/2,-a\,L^2/2\,\big)\, \Big(\frac{2}{a}\Big)^{1/2}+i\,\gamma\,\Big(5/2,-a\,L^2/2\,\Big)\, \Big(\frac{2}{a}\Big)^{5/2}\,\,\blue{\mathbf{\lambda}}\nonumber\\
&- \frac{i}{2}\,\gamma\,\big(9/2,-a\,L^2/2\,\big)\, \Big(\frac{2}{a}\Big)^{9/2}\,\,\blue{\mathbf{\lambda}^2} + \frac{i}{6}\,\gamma\,\big(13/2,-a\,L^2/2\,\big)\, \Big(\frac{2}{a}\Big)^{13/2}\,\,\blue{\mathbf{\lambda}^3}+\ldots
\label{series}
\end{align} 
We would now like to sum the infinite series \reff{ILS} and then take the limit as $L\to \infty$. As it stands, the sum over $n$ cannot be expressed in terms of any well-known function. This can be remedied by replacing the incomplete gamma function by its series representation \cite{Gradshteyn}
\beq
\gamma(\alpha,x)=\sum_{m=0}^{\infty} \frac{(-1)^m\,x^{\alpha+m}}{m! \,(\alpha+m)}\,. 
\eeq{Gam}
Substituting \reff{Gam} into \reff{ILS} we obtain 
\begin{align}
I_{series}(L)&=\sum_{m=0}^{\infty} \frac{a^m}{2^m\,m!} \sum_{n=0}^{\infty} \frac{(-1)^n}{n!} \,\lambda^n\,\frac{L^{4n+1 +2m}}{2n +1/2+m}\nonumber\\
&=\sum_{m=0}^{\infty} \frac{a^m}{2^m\,m!} \,\frac{1}{2}\, \lambda ^{-\frac{m}{2}-\frac{1}{4}}\, \gamma \,(\tfrac{m}{2}+\tfrac{1}{4},L^4 \lambda)
\label{Sum1}
\end{align} 
where the sum over $n$ was evaluated to     
\beq
\sum_{n=0}^{\infty} \frac{(-1)^n}{n!} \,\frac{L^{4n+1 +2m}}{2n +1/2+m}\,\lambda^n=\frac{1}{2} \,\lambda ^{-\frac{m}{2}-\frac{1}{4}} \,\gamma \,(\tfrac{m}{2}+\tfrac{1}{4},L^4 \lambda)\,.
\eeq{Sumn}
Since the original sum over $n$ of the series has been completed, we are now free to take the infinite $L$ limit in \reff{Sum1}. We obtain that 
\begin{align}
\lim_{L\to \infty} I_{series}(L)&=\sum_{m=0}^{\infty} \frac{a^m}{2^m\,m!} \,\frac{1}{2}\, \lambda ^{-\frac{m}{2}-\frac{1}{4}}\, \lim_{L\to \infty}\gamma \,(\tfrac{m}{2}+\tfrac{1}{4},L^4 \lambda)\nonumber\\
&=\sum_{m=0}^{\infty} \frac{a^m}{2^m\,m!} \,\frac{1}{2}\, \lambda ^{-\frac{m}{2}-\frac{1}{4}}\,\Gamma \,(\tfrac{m}{2}+\tfrac{1}{4})\nonumber\\
&=\frac{\pi}{4} \,  \sqrt{\frac{a}{\lambda }}\, e^{\frac{a^2}{32 \lambda }}\, \left(I_{\frac{1}{4}}\left(\frac{a^2}{32 \lambda }\right)+I_{-\frac{1}{4}}\left(\frac{a^2}{32 \lambda }\right)\right)\,.
\label{LF}
\end{align}
The final result in \reff{LF} is the analytical expression \reff{I2} of the original integral \reff{I1}. The infinite $L$ limit of the convergent perturbative series $I_{series}(L)$ has therefore reproduced the full non-perturbative result which corresponds here to the exact analytical expression of the integral.   

\subsection{Expansion about one of the two minima}

We will now obtain an expansion about one of the two minima but where the perturbative series is carried out under finite path integral limits. This will yield a convergent perturbative series in powers of $\lambda$. Recall that the original integral is given by  
\beq 
I= \int_{-\infty}^{\infty} e^{\frac{1}{2} a\,x^2 -\lambda \,x^4} \,dx
\eeq{IA1}
where $a>0$ and $\lambda>0$. 
As in section \ref{InfLimits}, we expand about the positive minimum $x_+=\sqrt{\tfrac{a}{4\,\lambda}}$ by writing  $x=\sqrt{\tfrac{a}{4\,\lambda}} +y$. The integrand, expressed in terms of the variable $y$, is expanded in a series about $\lambda=0$ which is equivalent to expanding the non-quadratic part about $y=0$. The infinite limits that appear on the integral over $y$ will now be expressed by writing finite limits $L$ on the integral with the limit of $L$ taken to infinity. This is an equivalent way of writing the integral. The only difference now is that the perturbative series expansion will be carried out under finite integral limits before the infinite $L$ limit is taken. The series expansion of integral \reff{IA1} about the minimum is given by 
\begin{align}
I_{series}&=e^{\tfrac{a^2}{16\,\lambda}} \int_{-\infty}^{\infty} e^{-a\, y^2 -2 \sqrt{a}\,\sqrt{\lambda }\, y^3-\lambda \, y^4} \,dy\nonumber\\
&=\lim_{L\to \infty} e^{\tfrac{a^2}{16\,\lambda}} \int_{-L}^{L} e^{-a\, y^2} \sum_{n=0}^{\infty}\frac{(-1)^n}{n!}\, (\,2\, \sqrt{a\,\lambda}\, y^3+\lambda \, y^4\,)^n\,dy\nonumber\\
&=\lim_{L\to \infty}\ I_{series}(L)
\label{IAS}
\end{align}
where 
\beq
I_{series}(L)=e^{\tfrac{a^2}{16\,\lambda}} \int_{-L}^{L} e^{-a\, y^2} \sum_{n=0}^{\infty}\frac{(-1)^n}{n!}\, (\,2\, \sqrt{a\,\lambda}\, y^3+\lambda \, y^4\,)^n\,dy\,.
\eeq{ISL2}
The above integral has limits running from $-L$ to $L$ where $L$ is finite, positive and real. Though $L$ is finite, it can be arbitrarily large.  We will develop a convergent perturbative series expansion of $I_{Series}(L)$ in powers of $\lambda$. The binomial expansion of the quantity inside the sum of \reff{ISL2} has already been worked out and is given by \reff{Binom}
\begin{align}
\frac{(-1)^n}{n!}\, (\,2\,\sqrt{a\,\lambda}\, y^3+\lambda \, y^4\,)^n\,
&=\sum_{k=0}^n (-1)^n\, \frac{2^{n-k}}{k!\, (n-k)!}\,\,a^{\tfrac{n-k}{2}}\,\,\lambda^{\tfrac{n+k}{2}}\,y^{3n+k}\,.
\label{BinomA}
\end{align}
The integral over $y$ runs from $-L$ to $L$ and it yields 
\beq
\int_{-L}^{L} e^{-a\, y^2}\,y^{3n+k}\, dy= \frac{1}{2}(1+(-1)^{n+k})\,a^{-\tfrac{(3n+1+k)}{2}}\,\gamma\big(\tfrac{3n+1+k}{2}, \,a \,L^2\big)
\eeq{Inty2}
where $\gamma\big(\tfrac{3n+1+k}{2},a \,L^2\big)$ is the incomplete gamma function defined by \reff{GG}. Substituting \reff{Inty2} and \reff{BinomA} into \reff{ISL2} yields
\beq
I_{series}(L)=e^{\tfrac{a^2}{16\,\lambda}}\,\sum_{n=0}^{\infty} \sum_{k=0}^n (-1)^n \frac{2^{n-k}\,a^{-n-k-\tfrac{1}{2}}}{k!\, (n-k)!}\,\,\tfrac{1}{2}\big(1+(-1)^{n+k}\,\big)\,\gamma\big(\tfrac{3n+1+k}{2}, a \,L^2 \big)\,\,\lambda^{\tfrac{n+k}{2}}\,.
\eeq{ISL3}
A simple way to view the effect of the finite $L$ is to note that the incomplete gamma function $\gamma\big(\tfrac{3n+1+k}{2}, a \,L^2 \big)$ in \reff{ISL3} has replaced the Euler gamma function $\Gamma\big(\tfrac{3n+1+k}{2}\big)$ in \reff{IS2}. In the sum \reff{ISL3}, only terms where $n+k$ is even are non-zero. Organizing the series order by order in integer powers of $\lambda$, we obtain the same result as \reff{IS3} except that $\Gamma\big(3q-k+1/2\big)$ is replaced by $\gamma\big(3q-k+1/2,a\,L^2\big)$, i.e.
\beq 
I_{series}(L)=e^{\tfrac{a^2}{16\,\lambda}}\,\sum_{q=0}^{\infty} \frac{4^q\,\lambda^q}{a^{2q+\tfrac{1}{2}}}\,\sum_{k=0}^q \frac{(-1)^k\,\gamma\big(3q-k+1/2,a\,L^2\big)}{4^k\,k!\, (2q-2k)!}\,.
\eeq{ISL4}
As it stands, the sum over $k$ does not yield any well-known function. However, by replacing $\gamma\big(3q-k+1/2,a\,L^2\big)$ by its integral definition \reff{GG}, we can sum over $k$ and then integrate. This yields 
\begin{align}
I_{series}(L)&=e^{\tfrac{a^2}{16\,\lambda}}\,\sum_{q=0}^{\infty} \frac{4^q\,\lambda^q}{a^{2q+\tfrac{1}{2}}}\,\int_0^{a\,L^2} dt \,e^{-t} \sum_{k=0}^q \frac{(-1)^k\,t^{3q -k-1/2}\,}{4^k\,k!\, (2q-2k)!}\nonumber\\
&=e^{\tfrac{a^2}{16\,\lambda}}\,\sum_{q=0}^{\infty} \frac{4^q\,\lambda^q}{a^{2q+\tfrac{1}{2}}}\,\int_0^{a\,L^2}dt\, e^{-t}\,\frac{t^{2 q-\frac{1}{2}} U\left(-q,\frac{1}{2},t\right)}{(2 q)!}\nonumber\\
&=e^{\tfrac{a^2}{16\,\lambda}}\,\sum_{q=0}^{\infty} \frac{\sqrt{\pi}\,4^q \,L^{4 q+1} \,\Gamma \left(2 q+\frac{1}{2}\right)\,\lambda ^q }{(2 q)!\,\Gamma \left(\frac{1}{2}-q\right) \Gamma(2 q+\frac{3}{2})} \, _2F_2\Big(q+\frac{1}{2},2 q+\frac{1}{2};\frac{1}{2},2 q+\frac{3}{2};-a L^2\Big) \nonumber\\
&=e^{\tfrac{a^2}{16\,\lambda}}\,\sum_{q=0}^{\infty} c_q\,\lambda^q
\label{La1}
\end{align}
where the coefficients $c_q$ are given by
\beq
c_q=\frac{\sqrt{\pi}\,4^q \,L^{4 q+1} \,\Gamma \left(2 q+\frac{1}{2}\right)}{(2 q)!\,\Gamma \left(\frac{1}{2}-q\right) \Gamma(2 q+\frac{3}{2})} \, _2F_2\Big(q+\frac{1}{2},2 q+\frac{1}{2};\frac{1}{2},2 q+\frac{3}{2};-a L^2\Big)
\eeq{cq}
and $_2F_2\left(q+\frac{1}{2},2 q+\frac{1}{2};\frac{1}{2},2 q+\frac{3}{2};-a L^2\right)$ is the generalized hypergeometric function\\ $_pF_q({a_1,..,a_p},{b_1,...,b_q}, z)$ with $p=2$, $q=2$ and $z=-a\,L^2$ and $U\left(-q,\frac{1}{2},t\right)$ is the confluent hypergeometric function. By the ratio test, the perturbative series \reff{La1} in powers of $\lambda$ is an absolutely convergent series for any finite $L$ since
\begin{align}
\lim_{q\to \infty} \Big|\frac{c_{q+1}\,\lambda^{q+1}}{c_{q}\,\lambda^{q}}\Big|=\lim_{q\to \infty}\,\Big|\frac{\lambda \,L^4 \,(4 q+1)}{(q+1)\,(4 q+5)}\frac{\, _2F_2\Big(q+\frac{3}{2},2 q+\frac{5}{2};\frac{1}{2},2 q+\frac{7}{2};-a L^2\Big)}{\, _2F_2\Big(q+\frac{1}{2},2 q+\frac{1}{2};\frac{1}{2},2 q+\frac{3}{2};-a L^2\Big)}\Big|=0
\label{Cqq}
\end{align}
where we used that the ratio of the two hypergeometric functions in the infinite $q$ limit yields unity.  As it stands, the infinite sum over $q$ cannot be carried out to yield any well-known function of $L$. However, this can be remedied if we express $_2F_2(a_1,a_2;b_1,b_2;z)$ in terms of its infinite power series. Then the sum over $q$ can be performed and we obtain  
\begin{align}  
I_{series}(L)&=e^{\tfrac{a^2}{16\,\lambda}}\sum_{m=0}^{\infty}\frac{\left(-a \,L^2\right)^m }{m! \,\Gamma \left(m+\frac{1}{2}\right)} \sum_{q=0}^{\infty}\frac{\pi\,4^q \,L^{4 q+1} \,}{(2 q)!\,\Gamma \left(\frac{1}{2}-q\right)} \, \frac{\Gamma \left(m+q+\frac{1}{2}\right) \Gamma \left(m+2 q+\frac{1}{2}\right)}{\Gamma \left(q+\frac{1}{2}\right)  \Gamma \left(m+2 q+\frac{3}{2}\right)}\,\lambda ^q \nonumber\\
&=e^{\tfrac{a^2}{16\,\lambda}}\sum_{m=0}^{\infty}\frac{2 L \left(-a \,L^2\right)^m} {(2 m+1)\, \Gamma (m+1)}\,\, _2F_2\Big(\frac{m}{2}+\frac{1}{4},m+\frac{1}{2};\frac{1}{2},\frac{m}{2}+\frac{5}{4};-L^4 \lambda \Big)\,.
\label{Sumq2}
\end{align}
Since the original infinite series over $q$ has been successfully summed to yield a function of $L$ (expressed as the sum \reff{Sumq2}), we are now free to take the infinite $L$ limit. This yields the exact analytical result of the original integral:
\begin{align}
&\lim_{L\to \infty} I_{Series}(L)\nonumber\\
&=e^{\tfrac{a^2}{16\,\lambda}}\sum_{m=0}^{\infty}\frac{2\, (-a)^m }{(2 m+1)\, \Gamma (m+1)}\lim_{L\to \infty} \Big[  L^{2m+1}\, _2F_2\Big(\frac{m}{2}+\frac{1}{4},m+\frac{1}{2};\frac{1}{2},\frac{m}{2}+\frac{5}{4};-L^4 \lambda \Big)\,\Big]\nonumber\\
&=e^{\tfrac{a^2}{16\,\lambda}}\sum_{m=0}^{\infty}\frac{2\, (-a)^m }{(2 m+1)\, \Gamma (m+1)}\,\Big[\,\frac{\sqrt{\pi } \,(2 m+1) \,\lambda ^{-\frac{m}{2}-\frac{1}{4}} \,\Gamma \left(\frac{m}{2}+\frac{1}{4}\right)^2}{4 \,\Gamma \left(\frac{1}{4}-\frac{m}{2}\right) \,\Gamma \left(m+\frac{1}{2}\right)}\,\Big]\nonumber\\
&=e^{\tfrac{a^2}{16\,\lambda}}\sum_{m=0}^{\infty}\frac{\sqrt{\pi } \,(-a)^m \,\lambda ^{-\frac{m}{2}-\frac{1}{4}} \,\Gamma \left(\frac{m}{2}+\frac{1}{4}\right)^2}{2 \,\Gamma \left(\frac{1}{4}-\frac{m}{2}\right) \,\Gamma \left(m+\frac{1}{2}\right) \Gamma (m+1)}\nonumber\\
&=\frac{\pi}{4} \,\, e^{\frac{a^2}{32 \lambda }} \,\sqrt{\frac{a}{\lambda }} \,\,\Big[\,I_{\frac{1}{4}}\big(\tfrac{a^2}{32 \lambda }\big)+I_{\frac{-1}{4}}\big(\tfrac{a^2}{32 \lambda }\big)\,\Big]\,.
\label{End}
\end{align}
The final expression in \reff{End} is the exact analytical expression \reff{I2} of the original integral \reff{I1}. The perturbative series about one of the minima has reproduced the full non-perturbative result for a system with a degenerate vacuum. This is a significant result as it implies that even when the the perturbative series is expanded about one of the minima it can still capture the full effects of both minima. We discuss this further in the conclusion.       

\section{Conclusion}

In this paper we showed that developing a perturbative expansion under finite path integral limits yields correct finite results even when the standard perturbative series is not Borel summable. Though we considered a basic integral, this possessed the two essential features of interest: it had a degenerate vacuum (i.e. a non-trivial vacuum structure) and its perturbative expansion led to an asymptotic series which was not Borel summable. The basic integral had an integrand $e^{-f(x)}$ where $f(x)=-a\,x^2/2 +\lambda \,x^4 $  with $\lambda>0$ and $a>0$. The function has a double-well shape containing two minima and a local maximum at the origin $x=0$. The first thing to note is that since $a>0$, it was not possible to establish a standard perturbative series about $x=0$. The reason is that the quadratic part $e^{a\,x^2/2}$ when integrated between infinite limits diverges exponentially (in contrast to the case $a<0$ which yields a Gaussian). It follows that an expansion of the quartic part in powers of the coupling $\lambda$ yields terms that are each infinite making it impossible to establish a series. One therefore had to carry out the series expansion about one of the minima (we chose the positive one at $x_+=\sqrt{a/(4\, \lambda)}\,$). We showed that this led to an asymptotic series which was not Borel summable since the Borel transform had a singularity on the positive real axis that lied along the path of integration of the Borel resummation.

When the perturbative expansion is carried out under finite path integral limits $L$, one can obtain a series by expanding about the origin $x=0$ or about one of the minima (we chose again the positive one at $x_+\,$). For both cases, we obtained a convergent perturbative series in powers of the coupling $\lambda$: the series \reff{ILS} for the case about $x=0$ and \reff{La1} for the case about the minimum $x_+$. After summing the series and then taking the infinite $L$ limit, we obtained the final expression 
\reff{LF} for the case about $x=0$ and the final expression \reff{End} for the case about the minimum $x_+$. In both cases, the final expression matched the exact analytical expression \reff{I2} of the original integral. This means the convergent perturbative series was able to capture the full effect of both minima. It has generally been thought that the effects that arise from a non-trivial vacuum, as in the case here, could not be fully captured by perturbation theory \cite{Zinn}. In particular, a perturbative expansion about one of the minima would appear to have little chance of capturing the `other side' of the double-well containing the other minimum i.e. of capturing the effect of both minima. This is reflected in the fact that standard perturbation theory yields an asymptotic series which is not Borel summable in such cases. However, under finite path integral limits, the perturbative expansion yields an absolutely convergent series and hence must reproduce exactly the original system regardless of its vacuum structure.       

One future goal is to extend this to other non-Borel summable cases such as the energy series of the double-well potential in quantum mechanics, which has been studied using multi-instanton methods \cite{Zinn2,Marino1}. Another case of considerable interest are the infrared (IR) renormalons that one encounters at large order in QCD and are not Borel summable \cite{Ioffe,Beneke} (the Borel resummation contains an uncertainty or ambiguity). Though QCD is significantly more challenging technically, the main reason for the Borel ambiguity appears to be the same: a non-trivial physical vacuum state \cite{Ioffe}. It is known that QCD also has UV renormalons but these are Borel summable \cite{Ioffe} (for recent work on Borel-improved perturbative expansions in QCD see \cite{Caprini}). So the main issue are not renormalons per say but that in the IR, the renormalons are associated with a non-trivial vacuum state. In this work, we have taken a first step in showing that finite path integral limits work when a system has a non-trivial vacuum and yields an asymptotic series which is not Borel summable. The next step is to apply this to more complex or realistic physical systems that also possess a non-trivial vacuum.    

\section*{Aknowledgments}
The author thanks Bishop's University for their financial support. The author thanks  the Physics Department at Princeton University for their invitation to present some preliminary work on this topic during April, 2026 which partly motivated this full study.  

\section*{Data Availability}
There are no publicly available data or software supporting this manuscript. Requests for further information or data should be sent to the author.

\end{document}